\documentclass[aps, amsfonts, nofootinbib, notitlepage, twocolumn, preprintnumbers]{revtex4-1}
\usepackage{hyperref}
\usepackage{ mathrsfs }
\hypersetup{
colorlinks=true,
linkcolor=red,
citecolor=blue,
urlcolor=blue}
\usepackage{epsf,epsfig}
\usepackage{amscd}
\usepackage{amsmath}
\usepackage{subfig}
\usepackage{graphicx}
\usepackage[countmax]{subfloat}
\newcommand{\bea}{\begin{eqnarray}}
\newcommand{\eea}{\end{eqnarray}}

\newcommand{\nn}{\nonumber}
\begin{document}
\title{Center symmetry and the Hagedorn spectrum} 
\author{Adi Armoni\footnote{{\tt a.armoni@swan.ac.uk}}} 
\affiliation{Department of Physics,\\
Swansea University, Singleton Park, Swansea, SA2 8PP, UK} 
\author{Thomas D.~Cohen\footnote{{\tt cohen@umd.edu}}} 
\author{Srimoyee Sen\footnote{{\tt srimoyee@umd.edu}}}
\affiliation{Maryland Center for Fundamental Physics,\\ 
Department of Physics,\\ 
University of Maryland, College Park, MD USA}

\begin{abstract}
	
This paper explores the  conjecture that large $N_c$ gauge theories have a Hagedorn spectrum, if, and only if, they are confining and posses an explicit or emergent center symmetry. Evidence in support of this conjecture is presented. Many classes of large $N_c$  gauge theories are considered. In all cases, we find that theories for which there exists a strong plausibility argument for a Hagedorn spectrum at large $N_c$ are also believed to be confining and possess either an explicit center symmetric or have a strong plausibility argument for the existence of an emergent center symmetry at large $N_c$.  Conversely,  all theories we considered which are believed not to have a Hagedorn spectrum at large $N_c$, either were known not to be confining or else were believed to lack an emergent center symmetry.  This is consistent with expectations based on the conjecture.
	
\end{abstract}
\maketitle

\section{Introduction}

The nature of confinement in non-Abelian gauge theories such as QCD remains an interesting and subtle problem\cite{Greensite:2011zz}. One clear intuitive notion of confinement is that all physically isolated states are color singlets. In this sense, QCD is a confining theory. However, a central difficulty in understanding the nature of confinement is that in QCD itself, and in many other theories, there is no order parameter for confinement. Other theories, such as pure SU($N_C$) Yang-Mills {\it do} have order parameters for confinement. It was pointed out long ago, by\cite{Polyakov:1978vu,Susskind:1979up} that the thermal expectation value of the Polyakov loop---a Wilson loop periodic in the time direction---acts as an order parameter.  Physically, it can be interpreted as 
 the free energy of a static color source by
\bea
\langle L \rangle=\exp(-\beta F)
\eea
where $L$ is the thermal expectation of the Polyakov loop, $\beta$ is inverse temperature and $F$ is the free energy. In the confined phase $F$ is $\infty$ and $\langle L \rangle=0$ and in the deconfined phase $F$ is finite and $\langle L \rangle$ is nonzero.    There has been a long history of studying the Polyakov loop as an indicator of the confining phase: The effective theory of Polyakov loops in $SU(2)$ Yang-Mills was shown to exhibit a second order phase transition just as in the $3-D$ Ising model by \cite{McLerran:1981pb}, \cite{Svetitsky:1982gs}. This was verified numerically by \cite{McLerran:1980pk,Kuti:1980gh,Engels:1980ty,Gavai:1982er,Gavai:1982fk,Engels:1989fz,Engels:1992fs} and the numerical estimation of the critical temperature was extrapolated to continuum \cite{Fingberg:1992ju}. $SU(3)$ Yang-Mills on the other hand was conjectured to have a first order phase transition \cite{Yaffe:1982qf} just as in the case of a three dimensional 3-state Potts model \cite{Wu:1982ra,Jensen:1979zz,herrmann,PhysRevLett.63.13,2336462}.   

The symmetry properties  of Polyakov loops play an essential role in gauge theories. Under center transformations---gauge transformations that are periodic in the time direction with periodicity $\beta$ up to a multiplicative factor of an element $z$ of the center $\mathbb{Z}_{N_c}$---the Polyakov loop transforms as $\langle L \rangle\rightarrow z\langle L \rangle$.   This implies that in a phase in which  $\langle L \rangle \ne 0$,  $\langle L \rangle$ changes under a center transformation, while in a phase in which  $\langle L \rangle = 0$,  it does not.  Thus, the Polyakov loop serves as an order parameter for center symmetry\cite{'tHooft:1977hy}.  Since it is also an order parameter for deconfinement in Yang-Mills theory, one sees that in Yang-Mills theory, the breaking of center symmetry is necessary for deconfinement.  

In QCD things are quite different: dynamical quarks act to screen static color charges and hence a static color source has a finite free energy and  $\langle L \rangle \ne 0$ at any temperature\cite{Greensite:2011zz}. Thus, in QCD the Polyakov loop is not an order parameter for confinement.  Moreover, the theory is not center symmetric, since center transformations spoil the boundary conditions for the quarks\cite{Makeenko}.

Although the notion of a center symmetric gauge theory in $2+1$ and $3+1$ dimensions is self -evident, it needs to be defined carefully in order to be applicable to $1+1$ dimensions as well. Note that, to conclude whether a theory is center symmetric or not we need an order parameter for the symmetry. In this case the order parameter is the Polyakov loop which is defined in a version of the theory which has compactified time direction. However, since the Polyakov loop is defined in the theory with compactified time direction, it tells us whether the compactified theory and not the theory in infinite space-time dimensions, is center symmetric or not. Hence, we need to define what we mean by a theory being center invariant in infinite space-time dimensions. The definition that we propose here has two parts. We define a theory on flat infinite space-time as possessing a center if two conditions are met: i) the theory when compactified has a center and ii) nontrivial correlation functions in the compactified theory have smooth infinite volume limits.

There is another indicator of confinement in Yang-Mills theory which is not present in QCD: an area law for a space-time Wilson loop, or equivalently a non-vanishing string tension at long distance.  While, the Wilson loop in infinite space time is neutral under center transformations, it has long been assumed that center symmetry is connected to confinement\cite{'tHooft:1977hy}. Recently it has been shown that the connection is quite strong: theories in a very large class---all SU($N_c$) gauge theories in 3+1 or 2+1 space-time dimensions with possible matter fields in any representation as well as theories in 1+1 dimensions containing matter fields in the adjoint plus possible other representations---have been shown to have an area law for the Wilson loop only if the theory is invariant under a nontrivial subgroup of the  $\mathbb{Z}_{N_c}$ center \cite{Cohen:2014swa}.  

Any attempt to understand the role played by center symmetry in  confinement depends sensitively on precisely what one means by ``confinement''.   If one equates confinement to a theory having properties such as a vanishing Polyakov loop or an area law for a Wilson loop, then center symmetry plays a critical role.  Such a perspective is advocated in ref.~\cite{Greensite:2011zz}.  While, this is a reasonable way to proceed since properties such as these are amenable to theoretical attack, it has the disadvantage that QCD---the physical theory which gave rise to the notion of confinement---cannot be described as confining but only as a``confinement-like'' theory.   In this paper, we take ``confinement'' to mean color confinement---the property that all isolated physical particle states in the theory are color singlets. Note that to be confining with the current definition, two features are needed: i) isolated color neutral particle states must exist in the theory and ii) no colored physical  particle states exist.  Condition i) is clearly important in this as well as condition ii).  Thus, for example, conformal gauge theories such as N=4 SYM are not confining with this definition since it does not have a particle spectrum and hence lacks color neutral particles states.   With this definition, center symmetry is clearly not required for confinement---QCD is confining in this sense, yet lacks center symmetry.  What role, then, does center symmetry play?  The perspective taken here is that while confinement itself does not require center symmetry, certain indicators of confinement---such as a vanishing Polyakov loop---require both confinement itself {\it and} center symmetry.

The present paper is aimed at understanding the role of center symmetry in a somewhat different kind of an indicator of confinement---namely, the existence of a Hagedorn spectrum. A convenient way to parameterize the hadronic spectrum is in terms of $N(m)$, a function which gives the  number of hadrons with mass less than $m$. A Hagedorn spectrum is one which at large masses scales as 
\begin{equation}
N(m)\sim f\left(\frac{m}{T_H}\right) \exp(\frac{m}{T_H})
\end{equation}
where $T_H$ is a parameter known as the Hagedorn temperature and $f$ is a function which grows more slowly than an exponential. This type of spectrum was originally conjectured by Hagedorn in 1965 \cite{hagedorn1,hagedorn2} on phenomenological grounds.

Before discussing the sense in which a Hagedorn spectrum is an indicator of confinement, it is useful to discuss briefly the status of the Hagedorn conjecture of an exponentially growing spectrum of hadrons.  The form of the subexponential  prefactor---often taken to be a power law at large mass---plays a significant role in obtaining a reasonable fit  with empirical data for hadron masses over a large range of masses\cite{PhysRevD.70.117503}.  Generally,  the data from extracted masses of hadronic resonances are qualitatively consistent with the existence of a Hagedorn spectrum up through a bit over 2 GeV.  However, by themselves these data are insufficient to give compelling evidence for Hagedorn's conjecture \cite{Cohen:2011cr}. As a practical matter, one difficulty is that a compelling evidence for a Hagedorn spectrum would require data through large enough masses to be able to distinguish exponential behavior from that dominated by the subexponential prefactor.  However, the hadronic resonances are very difficult to find for masses above around 2.3 GeV (for the case of hadrons containing no heavy quarks) and there is not enough of a dynamic range to distinguish between exponential from subexponential growth in a compelling way.  Of course, if this inability to find very high mass resonances is because these resonances do not exist, then the Hagedorn conjecture that at high masses the number of hadrons grows exponentially is simply incorrect.  However, it is quite plausible that such resonances do exist but as the mass increases it becomes increasingly difficult to distinguish such resonances in the data. Note that the number of open decay channels grows with mass as does the phase space for the decay of each resonance making it exceptionally difficult to find compelling evidence for resonances in partial wave analysis from the scattering data, even if  resonances do exist.  

It is worth noting, at this stage, that the difficulty of identifying high lying resonances has theoretical as well as practical implications. After all, the Hagedorn conjecture relates the number of states to the mass of the hadron.  However, the mass is not really a well-defined quantity for resonances since the resonances have widths. It is precisely the widths  of the resonances that render difficult the problem of identifying high lying resonances. The extraction of resonances from scattering processes of various kinds necessarily involves some level of model dependence. When resonances are well separated and narrow, this model dependence is rather modest and as a practical matter one can test to see the extent to which the growth in the number of hadrons appears to be exponential. However, as the masses increase one has wide and overlapping resonances and there is considerable model dependence, leading to the practical issue in testing the conjecture.  To the extent, that the masses of hadrons are not  well-defined there is a fundamental ambiguity in the formulation of the Hagedorn conjecture.

Let us return to the issue of the sense in which a Hagedorn spectrum is associated with confinement.  At a  crude level it is not surprising that a Hagedorn spectrum could be connected with confinement.  There is a very natural picture of confinement as being associated with flux tubes which act like strings for sufficiently long flux tubes that the width of the tubes becomes irrelevant\cite{Nambu:1978bd}. Pure gauge theories do act stringy. At a simple level, the potential between widely separated color sources rise linearly with distance \cite{rothe}. This is because the gluon fields between two static color sources arrange themselves into tubes with  a fixed energy per unit length. It is natural to expect that highly excited hadrons having long and thin flux tubes would be described by a theory that is very much like string theory. In this picture open strings are interpreted as mesons while  closed strings are interpreted as glueballs. In a flux-tube picture, highly excited hadrons are associated with long flux tubes and with only small corrections to  act like a string. Now, as it happens one of the earliest things known about string theories is that  simple string theories give rise to Hagedorn spectra \cite{polin}.  

However, there is something wrong with this picture when applied to a QCD string.   Hagedorn spectra do arise in simple string theories; but simple in this context means non-interacting and unbreakable.  However, real QCD strings both interact and break. This fact means that the stringy explanation of a Hagedorn spectrum is at best incomplete.   This should come as no surprise, this is the stringy incarnation of the issue noted above; the fact that hadrons are resonances rather than bound states and accordingly have widths. Indeed, the fact that strings break is what allows mesons to decay into two mesons; the fact that strings self-interact  allows glueballs  to ``bud off'' additional glueballs and decay. It is worth noting that this issue is not directly associated with the lack of center symmetry in QCD.  Center symmetric theories, such as pure Yang-Mills theory, while lacking string breaking, do have string-string interaction which allow glueballs to decay and thereby create widths for glueballs. 

At a theoretical level, these problems can be avoided---but at a cost. The cost is that rather look directly at QCD (or some other gauge theory) itself. One can look at the large $N_c$ limit  of the theory \cite{'tHooft:1973jz,Witten:1979kh}. Note that in this limit flux tubes do not break\cite{PhysRevD.20.179}. So a description in terms of a simple string theory (with a Hagedorn spectrum) is plausible. Moreover, meson and glueball decays are suppressed by powers of $N_c^{-1}$ \cite{Witten:1979kh}. Thus, the masses of hadrons become  well defined in this limit as their widths go to zero. This means that the question of whether or not a Hagedorn spectrum exists becomes well posed. Note that there is an implicit issue with the ordering of limits here. In a strict sense the Hagedorn spectrum implies exponential growth asymptotically high in the spectrum. One needs to take the large Nc limit prior to going asymptotically high in the spectrum.

While the existence of a Hagedorn spectrum at large $N_c$ is both a well-posed question and is highly plausible, a key question is whether it actually exists.  One way to get insight into the problem is via studies in 1+1 space-time dimensions.    There is a long history of studying the spectra of large $N_c$ QCD in 1+1 dimensions. On one hand,  in the case in which there are {\it massive} quarks in the adjoint, as opposed to the fundamental representation, there have long been arguments that a Hagedorn spectrum should emerge \cite{Kogan:1995nd,Bhanot:1993xp,Kutasov:1993gq,Demeterfi:1993rs,Gross:1997mx}. On the other hand, the spectrum for large $N_c$ QCD in 1+1 dimension with quarks in the fundamental representation has been known since 't Hooft's seminal work\cite{'tHooft:1977hy}; the spectrum of this theory is most emphatically {\it not} of the Hagedorn sort: rather than growing exponentially with particle mass, $N(m)$ grows quadratically with $m$.

What is the origin of the difference in behaviors of these two theories in 1+1 dimensions ?  Both of these theories are confining.  However, there is an obvious qualitative difference between the two theories:  The theory with massive quarks in the adjoint representation, is center symmetric, while the one with quarks in the fundamental representation is not. It seems plausible that the qualitative differences between the large $N_c$ spectra of the two theories is tied to center symmetry.  The purpose of the present paper is to explore the conjecture that the existence of a Hagedorn spectrum at large $N_c$ is an indicator of confinement for a theory, but like some other indicators of confinement such as the area-law for Wilson loops, it only becomes manifest for theories which are center symmetric as well as confining.

The conjecture is not about a gauge theory itself. Rather, it is a conjecture about the large $N_c$ limit of the theory.  
 However, there are a variety of ways that a large $N_c$ limit can be taken.  Traditionally, the large $N_c$ limit is taken in the manner of `t Hooft---$N_c \rightarrow \infty$, $g \rightarrow 0$ with $g^2 N_c$ held fixed.  The conjecture applies to theories in this standard limit.  However, there exist  theories with fermions in higher representations, {\it i.e.} representations whose dimension scales with $N_c$ as $N_c^a$ with $a>2$ at large $N_c$. While such theories are problematic in the ultraviolet in 3+1 space-time dimensions, they are perfectly sensible in 2+1 or 1+1 dimensions. Such theories have been studied recently \cite{Cohen:2014gka,Poppitz:2009kz}, and while a consistent large $N_c$ limit can be constructed for such theories, it is distinct from the usual `t Hooft limit. In particular, to be sensible, these theories require the limit be taken with  $N_c \rightarrow \infty$, $g \rightarrow 0$ with $g^2 N_c^{a-1}$ held fixed.  The conjecture discussed in this paper is not expected to apply to these nontraditional  $N_c$ limits. Moreover, the conjecture is also limited to theories which are ultraviolet complete.  This restriction is for an obvious reason: Hagedorn behavior, strictly speaking applies to the spectrum in the limit that the particle masses become large---that is in the UV part of the spectrum. This condition, along with the traditional `t Hooft limit restricts the matter content of the theories for which  the conjecture applies to either  pure Yang-Mills theories with no matter content or to  theories in which the  matter is restricted to being in the fundamental representation, the adjoint representation, the two-index symmetric representation or the two-index antisymmetric representation.
 
Obviously, the fact that the conjecture only applies to the large $N_c$ limit, implies that it is restricted  to theories where the gauge group, {\it has} an $N_c$ which can be taken large, {\it i.e.} SU($N_c$), SO($N_c$) and Sp($N_c$) ) groups. Thus, for example, it is meaningless to ask whether an exceptional group such as G2, which has received a lot of attention over the years\cite{Holland:2003jy,Maas:2012ts,Cossu:2007dk}, has a Hagedorn spectrum at large $N_c$.

There is a subtlety associated with the  fact that the conjecture applies at large $N_c$. One can easily envision circumstances in which a class of theories at any finite $N_c$ is not center symmetric but for which an emergent center symmetry arises dynamically in the large $N_c$ limit.  Since the conjecture is not for a theory itself but rather for the large $N_c$ limit of the theory, it is highly plausible that the conjecture, if correct, should apply in such cases and imply that such theories, at large $N_c$, will have Hagedorn spectra if confining.  Of course, to make this concrete one needs to specify what one means by an emergent center symmetry. This issue was discussed in some detail in ref.~\cite{PhysRevD.77.045012}. In this paper, we will define an emergent center symmetry of a theory without an explicit center in a slightly different, although essentially equivalent way. For our purposes, an emergent center symmetry of a theory without an explicit center will be taken to exist provided that there exists a sector of  operators in the theory with nontrivial correlation functions that match, up to corrections which vanish as $1/N_c$, the correlation functions of a distinct theory which is explicitly invariant under center transformations. That is, the two theories are equivalent up to 1/$N_c$ corrections within a common sector. The emergent center symmetry can be identified as the explicit center of the equivalent theory.  

There is one additional issue that arises when one tries to relate center symmetry to spectral properties of the hadrons in the theory. We are interested in the hadronic spectrum as seen in the flat space  zero temperature and infinite volume limit of the theory.  However, strictly speaking to define an order parameter for center transformations we need to study a space which is compact in at least one dimension. We do this by going to Euclidean space with a finite temporal extent  of $\beta$ and  periodic boundary conditions for bosons and anti-periodic fermions; this is, of course the thermal theory with $\beta=1/T$.  Since we wish to consider the zero temperature limit to look at the spectrum, we generally identify a theory as being center symmetric if the finite temperature version of the theory is invariant under transformations and retains this property as the zero temperature limit is taken.  For most cases this is quite innocuous, but for SU($N_c$) Yang-Mills theories in 1+1 dimension, the theory at $\beta=\infty$ is qualitatively different from at finite $\beta$.   

The conjecture in its strongest form is that any gauge theory based on SU($N_c$), SO($N_c$) or Sp($N_c$) group and any matter content will have a Hagedorn spectrum in the `t Hooft large $N_c$ limit, if, and only if, the theory satisfies the following three conditions: i) the large $N_c$ limit exists, ii) the theory is confining (in the sense discussed above) and iii) the theory possesses an exact or emergent center symmetry (in the sense discussed above).

While a mathematically rigorous proof of this conjecture is beyond the scope of this paper, it will be argued here that the conjecture is highly plausible. Ideally, it could be determined conclusively whether theories in a particular class  satisfy the conditions of the conjecture and whether they have a Hagedorn spectrum. Then the conjecture would be established if it could be shown on a case-by-case basis that every class of theories that satisfies the assumptions has a Hagedorn spectrum and  that no theory which does not satisfy them has a Hagedorn spectrum. In principle, it might be possible to do this, since even though there are an infinite number of possible theories, one might hope to establish this for entire classes of theories.    Unfortunately  this is beyond the current state of the art.  We do not know how to definitively establish in a mathematically rigorous way whether or not an emergent center symmetry exists at large $N_c$ in cases where there is no explicit center. Similarly, we do not know how to definitively establish in a mathematically rigorous way whether or not a Hagedorn spectrum exists.

However, there is a plausibility argument that certain theories have Hagedorn spectra at large $N_c$\cite{Cohen:2011yx}. Similarly, there are strong plausibility arguments that various theories have emergent center symmetries\cite{PhysRevD.77.045012}. What we can show is that every class of theories we have, for which the plausibility argument for a Hagedorn spectrum applies either is explicitly center symmetric or there is a strong plausibility argument for an emergent center symmetry. Similarly, every theory we have considered that is known not to have a Hagedorn spectrum in large $N_c$ is either not confining or is not center symmetric and for which there is no reason to believe that a center symmetry  emerges at large $N_c$. While this is clearly not sufficient to prove the conjecture, it does provide significant evidence in its favor.

This paper is organized as follows: In the next section, theories which are known not have Hagedorn spectra at large $N_c$ are discussed. Consistent with the conjecture these are theories which are either not confining or lack any evidence for an emergent center. As it happens, there are relatively few of these. In the following section the plausibility argument for the existence of a Hagedorn spectrum is reviewed briefly. Following this 

\section{Large $N_c$ gauge theories without Hagedorn spectra \label{wo}}

As will be seen in the following sections, a very large class of gauge theories appear to both have Hagedorn spectra at large $N_c$ and either explicit or emergent center symmetries. However, there are a few known exceptions. These are instructive as the ones that lack a Hagedorn spectrum appear to lack either confinement or  center symmetry.  In this section these cases will be discussed.

\subsection{The 't Hooft model}

The 't Hooft model \cite{'tHooft:1973jz} is large $N_c$, SU($N_c$) in 1+1 space time dimensions with one flavor of quark which is in the fundamental representation. This theory is known not to have a Hagedorn spectrum. It is easy to see this. The spectrum at large $N_c$ consists of mesons which are arbitrarily narrow and arbitrarily weakly interacting. The meson spectrum is calculable in terms of an integral equation which, in general needs to be solved numerically. However, the equation becomes  trivial to solve for asymptotically high excited states. The masses of the highly excited mesons in this theory are given by 
\begin{equation}
m_n^2 = g^2 \pi n \label{HooftModel}
\end{equation}
where $n$ are (large ) positive integers and $ g^2$ is the (dimensionful) coupling constant of the  1+1 dimensional theory. Note, that the energy levels given in Eq.~ (\ref{HooftModel}) are non degenerate. The upshot of this is that for the 't Hooft model
\begin{equation}
N(m) \sim m^2
\end{equation}
which is manifestly slower than the exponential spectrum required for a Hagedorn spectrum.

Theories in more than 1+1 dimensions whose matter content is  a  single flavor of quark in the fundamental representation (or indeed, any fixed finite number of quarks in the fundamental) have an  emergent center symmetry in the large $N_c$ limit.    The reason is simple: quark loops are suppressed at large $N_c$ if the quarks are in the fundamental representation.  Thus, the gluodynamics at large $N_c$ does not depend on the quarks; accordingly as $N_c \rightarrow \infty$ correlation functions become those of pure Yang-Mills theory.  Since pure Yang-Mills theory is center symmetric, such theories have an emergent center symmetry, according to the definition given in the  introduction: that there exists a sector of operators in the theory with nontrivial correlation functions that match, up to corrections which vanish as $1/N_c$, the correlation functions of a distinct theory which is explicitly invariant under center transformations.  However, this argument does not apply for 1+1 dimensions.  In 1+1 dimensions  in flat infinite space, gluons are  not dynamical; their only role is to mediate interactions between the quarks.  Thus, unlike in higher dimensions,  pure Yang-Mills theory in 1+1 in infinite flat space-time does not exist as a dynamical theory.   Thus, the argument that theories with only quarks in the fundamental representation have emergent center symmetries  breaks down in 1+1 dimensions---there is no common sector with nontrivial correlation functions since the Yang-Mills theory has none.  Of course, absence of proof is not the same as proof of absence, and the breakdown of the argument is not logically equivalent to a proof that a center symmetry does {\it not} emerge at large $N_c$.  One might imagine one appearing for some  unrelated, and totally unknown, reason.  However, this seems extremely unlikely. 

Thus, the `t Hooft model is one which apparently neither has a Hagedorn spectrum nor an emergent center symmetry.  This is consistent with the conjecture.

There is, however, a subtlety associated with the preceding conclusion. As noted in the introduction to probe whether a theory has a center symmetry one considers the theory on a domain which is finite in some direction---for concreteness we consider the Euclidean theory with finite temporal extent to obtain the thermal theory.  And while the pure Yang-Mills theory in Euclidean space and infinite temporal extent {\it is} trivial, the theory with finite spatial extent is not: nontrivial spatial correlators exist for Polyakov loops that wrap around in the temporal direction.  Moreover, these same correlators could be computed in the `t Hooft model with finite temporal extent and, in the large $N_c$ limit they would match those of Yang-Mills---a theory which is center symmetric.  Thus, the question of whether or not the theory has an emergent center, appears to  depend on how an ordering of limits is applied to the definition.  One needs to take the  $N_c \rightarrow  \infty$ limit to establish the connection between the `t Hooft model and Yang-Mills and the $\beta \rightarrow \infty$ limit to go from a regime where the center is defined to the regime of interest for the spectroscopy.  If one were to start at finite but large $N_c$ and $\beta$ and first take the $N_c \rightarrow  \infty$ limit--connecting the two theories in a regime with nontrivial correlators and subsequently the $\beta \rightarrow \infty$ limit one might be tempted to conclude that an emergent center exists.  In contrast, if one first takes the $\beta \rightarrow \infty$ limit  and then the $N_c \rightarrow  \infty$,  the two theories share no nontrivial correlators at large $N_c$ and one concludes that an emergent center does not exist.  We believe that in the present context, the second ordering is the sensible one.  We note that for the present purpose, the reason that spectral functions are of interest was solely because they are connected to the hadron spectrum.  However, in the first ordering the only correlators that are matched are ones which have no connection to the hadronic spectrum and indeed ones which become ill-defined for infinite $\beta$.   Given this, we believe that the second ordering is justified, and with it the example remains consistent with the conjecture.

The situation is analogous for a somewhat more general class of theories.  The case considered above was for one flavor of quark in the fundamental representation. What happens if there are more? Note that the $n^{\rm th}$ state is not degenerate for the one flavor variant of the theory; in general, it has a degeneracy given by $N_f^2$ regardless of $n$ for the  case of $N_f$ degenerate flavors.  Note that provided $N_f$ is finite, even if the masses are unequal one will to a very good approximation have a degeneracy of $N_f^2$ with masses given by Eq.~(\ref{HooftModel}) if $n  \pi g^2 \gg m_h^2$ where $m_h$ is the mass of the heaviest quark.  Thus, again $N(m) \sim m^2$, and there is no Hagedorn spectrum.  Similarly, in the multi-flavor case there is still no reason to believe that a center symmetry emerges.

\subsection{1+1 QCD with massless adjoint quarks}\label{massless}

Two dimensional QCD with adjoint quarks is a subtle case. Whether the theory exhibits a Hagedorn spectrum or not depends on whether the quarks are massless or massive. The story is as follows: in the mid 70's Coleman et.al. showed that 1+1 QED with massless electrons is always screening \cite{Coleman:1975pw}, namely that massless electrons can screen fractional charges. The non-Abelian analogue of this fascinating phenomenon had been conjectured by Gross et.al. in the mid 90's \cite{Gross:1995bp}. Gross et.al. argued that massless adjoint quarks can screen fundamental quarks. A semiclassical calculation of the string tension in 1+1 QCD had been carried out in \cite{Armoni:1997ki} with the result that the string tension is proportional to the quark mass, $\sigma \sim m_q g$ (where $m_q$ is the quark mass and $g$ is the gauge coupling). This is distinct from the four dimensional case, where the QCD string tension is not expected to vanish as the quark mass goes to zero.

The vanishing of the string tension in 1+1 QCD with massless quarks has dramatic implications on the spectrum. The spectrum is not expected to exhibit a Hagedorn behavior \cite{Gross:1997mx}. A particular case where this is seen clearly, is when the number of flavors is large $N_f\gg 1$. The theory is becoming effectively Abelian in this limit and the spectrum consists of a single massive meson of mass $m^2 = {g^2 N_c N_f \over \pi}$  \cite{Armoni:1995tf}, as in the massless Schwinger model.

Naively, is seems that the absence of Hagedorn behavior in the present case contradicts our claim about confining theories with a center symmetry. Let us argue now that there is no contradiction: while the center symmetry is present at the classical level, it gets broken spontaneously. This is linked to the vanishing of the string tension.

The expectation value of a Polyakov loop is closely related to the free energy of an isolated quark. In the confining phase this energy is infinite and $\langle L \rangle =0$, while in the deconfining phase this energy is finite and  $\langle L \rangle \ne 0$
\begin{equation}
|\langle L \rangle|  \sim \lim _{L\rightarrow \infty} \exp ( - \beta \sigma L)   \, ,
 \end{equation}
 where $\beta$ is the compactification radius. A theory with a vanishing string tension is always in the deconfining phase, with $\langle L \rangle \ne 0$, and thus the center of the gauge group is spontaneously broken.

\subsection{The Veneziano limit}

The Veneziano limit of QCD is one in which the the number of flavors of quarks (which are in the fundamental representation) goes to infinity as $N_c$ goes to infinity with the ratio $N_f/N_c$ held fixed \cite{Veneziano:1979ec}. Theories approaching the limit are manifestly ultraviolet complete in dimensions less than 3+1; in 3+1, the theory is asymptotically free (and hence  ultraviolet complete) when $N_f/N_c < \frac{11}{2} $.  

The key reason that the Veneziano limit is interesting concerns the width of hadrons.  In the usual `t Hooft large $N_c$ limit with quarks in the fundamental representation, mesons and glueballs are both parametrically narrow, scaling as $1/N_c$ and $1/N_c^2$ respectively. This is significant, from the perspective of the Hagedorn spectrum. As noted in the introduction, it is only because hadrons remain narrow at large $N_c$ that the concept of a Hagedorn spectrum becomes truly well-defined in a mathematical sense at large $N_c$. However, in the Veneziano limit, neither the mesons nor the glueballs are parametrically narrow. Note that generically when one includes some number of active flavors $N_f$, the meson and the glueball widths scale as $N_f/N_c$ and $(N_f/N_c)^2$ respectively. In the Veneziano limit these are of order unity.  Since the widths are not in any sense parametrically narrow, the hadron masses are not well-defined even as the the large $N_c$ limit is taken. Accordingly there is no Hagedorn spectrum.

It is also noteworthy that the argument for an emergent center symmetry at large $N_c$ based on quark loops becoming suppressed breaks down in the Veneziano limit. The reason is again simple: The suppression factor for quark loops is parametrically $N_f/N_c$. For any fixed $N_f$, the loops become suppressed as $N_c \rightarrow \infty $ and the explicitly center-symmetric gluodynamics decouples from the quark dynamics yielding an emergent center symmetry. However in the Veneziano limit, the quark loops remain of order unity and quark dynamics do not decouple at large $N_c$ and thus center symmetry does not emerge in this limit. Of course, again this is not logically equivalent to a proof that a center symmetry does {\it not} emerge at large $N_c$. However, again it seems extremely unlikely. Thus, the Veneziano limit again appears to be a case in which there is no Hagedorn spectrum at large $N_c$ and no emergent center symmetry.

\subsection{Conformal Theories}

If a theory has a Hagedorn spectrum, i.e. a spectrum of particles whose number grows exponentially with the mass, the theory must, at the very least have particles. However, theories which are conformal, such as ${\cal N}=4$ SYM \cite{Maldacena:1997re} or which flow to conformal in the infrared (i.e. theories whose gauge coupling has an infrared fixed point) such as three index quarks in less than four dimensions \cite{Cohen:2014gka}, do not have particles and thus clearly do not have Hagedorn spectra. Conformal theories are gapless and two-point  correlation functions in such theories fall off at long Euclidean distances as power laws as opposed to exponentials.  However, these theories are also not confining as the term is commonly understood  The usual notion of confinement is that isolated physical particles exist and must be colorless. In conformal theories, there are no isolated colorless particles, for the simple reason that there are no particles. Thus, theories are typically divided between those which are confining and those which are conformal. Indeed, one of the major challenges in lattice gauge theory is to determine which theories are confining and which are conformal\cite{Svetitsky:2013px}.  

Thus we see that theories which flow to a conformal fixed point in the IR lack a Hagedorn spectrum but also lack confinement.

In this section we have presented a few types of theories which lack either an emergent center symmetry or confinement; all of these theories also lack a Hagedorn spectrum at large $N_c$. We know of no theory which lacks either an emergent center or confinement and also has a Hagedorn spectrum.

\section{Conditions for a Hagedorn Spectrum \label{Hagcond}}

In the previous section, it was shown that theories that are believed to lack either an emergent center symmetry or confinement also lack a Hagedorn spectrum. In the remainder of the paper we provide evidence for the converse: namely that theories which are both confining and have either an explicit or an emergent center symmetry at large $N_c$ also have a Hagedorn spectrum in the large $N_c$ limit.

The evidence is incomplete. Ideally, to prove that a theory has a Hagedorn spectrum one would calculate the spectrum of hadrons at large $N_c$ and demonstrate that the number grows exponentially with the mass. However, we do not have a viable way to calculate explicitly the  masses of high-lying hadrons at large $N_c$ in any theory except the `t Hooft model.  Note that  Lattice QCD is not particularly helpful either for the extraction of high-lying states. Transverse lattice QCD in light -cone formalism was used to study large $N$ glueball spectrum \cite{Dalley:2004ca}, the results of which were consistent with the existence of a Hagedorn spectrum but cannot be considered as definitive in any sense. Alternatively, one might hope to be able to constrain the number of states from below in  a mathematically rigorous manner,  and show that the number grows exponentially even if one cannot compute the energy levels of specific states.  Unfortunately, that too is beyond the state of the art.  

However, there does exist an argument for the existence of a Hagedorn spectrum for particular theories. This argument depends on a critical assumption that is  not  mathematically rigorous, but is highly plausible. If one accepts this argument, then one can deduce the existence if a Hagedorn spectrum in a very broad class of theories. In this paper we  closely follow the analysis  of this argument  in \cite{Cohen:2011yx} and \cite{Cohen:2009wq}.   The approach has similarities with some prior work\cite{Kogan:hep-ph9509322,Sundborg:1999ue,Aharony:2003sx,Aharony:2005bq}.  The critical assumption  underlying the argument is that correlation functions of composite operators are accurately described at short distances by their asymptotically free results plus corrections due to renormalization group-improved perturbation theory whenever the correlation functions are in the regime that renormalization group-improved perturbation theory indicates that the corrections to the asymptotically free results are small. In effect, the requirement is that RG-improved perturbation theory can be used reliably to estimate the regime of validity of perturbation theory itself. While this assumption is certainly not rigorous from a mathematical perspective, it is commonly assumed in the field.  Indeed,  asymptotic freedom of QCD has only been demonstrated in QCD\cite{PhysRevLett.30.1343} up to the validity of this assumption.  The argument also depends on the theory being confining in its most basic sense of the term---that is, that  all isolated states are color singlets.  It is noteworthy that confinement in this sense is quite distinct from  an unbroken center symmetry; and an unbroken center  is not invoked in the argument. More generally, stringy dynamics is not explicitly assumed.

Assuming that the assumptions underlying the argument of references \cite{Cohen:2011yx} are correct, a theory will have a Hagedorn spectrum if the following conditions are satisfied:
\begin{enumerate}
\item  Hadrons with zero baryon number ({\it eg.}  mesons) have widths that go to zero as $N_c \rightarrow \infty$ and in that limit single hadron states saturate the spectral function of the  Kallen-Lehman representation for  correlation functions of single-color-trace operators carrying zero baryon number  . \label{cond1}
\item The number of  local gauge-invariant  linearly-independent  single-color-trace operators with some fixed quantum numbers in the theory grows exponentially with the naive mass dimension of the operator: there exists a constant, $a$, such that for  mass dimension, $d$, beyond some minimum, the number of independent operators with the specified quantum numbers and naive mass dimension $d$ or less,  is larger than  $e^{a d}$.\label{cond2}
\item The theory is asymptotically free\label{cond3}.
\item  In the large $N_c$ limit, the corrections at any order in renormalization-group-improved perturbation theory at short distance to the free-field result for the derivative with respect to separation in Euclidean space of the  logarithm of the matrix of two-point correlation function for a set of single-color-trace operator that grows exponentially with the dimension of the operators   have the property that their fractional size asymptotically (in the dimension of the operators in the set) does not grow with the dimension of the operators.\label{cond4}
\end{enumerate}

Condition \ref{cond4} is at the crux of the argument for the existence of a Hagedorn spectrum.  The demonstration in ref.~(\cite{Cohen:2011yx})  that this condition is satisfied  for large $N_c$ QCD is somewhat intricate.  It was formulated explicitly for $SU(N_c)$ gauge theories containing quarks in the fundamental representation.  However, the critical issue in the demonstration is that the dominant graphs at large $N_c$ are planar.  The dominance of planar graph at large $N_c$ is true for pure Yang-Mills theory and for theories with quarks in the adjoint-representation, the two-index symmetric representation or the two-index anti-symmetric representation as well as for gauge theories  based on SO($N_c$) and Sp($N_c$). As it happens, the demonstration in ref.~(\cite{Cohen:2011yx}) holds without substantial modification for all of these cases.

Note that the examples of theories without Hagedorn spectra discussed in the previous section do not satisfy one or more of these conditions.  Thus for example, in the 't Hooft model, condition \ref{cond2} is not satisfied: one cannot construct a set of single-color trace operators which grows with the dimensions. Similarly for both the Veneziano limit and conformal theories condition \ref{cond1} is not satisfied.

\section{Theories with Hagedron Spectra}

In this section we look at different theories at large $N_c$ starting with  SU($N_c$) theories including pure Yang-Mills, and theories with quarks in the fundamental representation, and quarks in the adjoint in various space-time dimensions in order to investigate the existence of center symmetry and Hagedorn spectrum. We also consider theories with multiple fundamental indices on quark fields and SO($N_c$) and Sp($N_c$) gauge theories.  

\subsection{SU($N_c$) gauge theories with quarks in the fundamental representation} 

We begin with a brief discussion of gauge theories with quarks in the fundamental representation in various space-time dimensions. This discussion can be brief as this case was discussed extensively elsewhere \cite{Cohen:2011yx}. As noted in Sect.~\ref{wo}, in 1+1 dimensions this is the `t Hooft model and has neither an emergent center symmetry nor a Hagedorn spectrum. The cases of $2+1$ and  $3+1$ space-time dimensions were shown to have Hagedorn spectra in the large $N_c$ limit given the assumptions discussed above \cite{Cohen:2011yx}. The reader is referred to that work for details.  These theories do not have an explicit center symmetry for any finite $N_c$ due to the presence of the quarks. However, at large $N_c$ the gluodynamics is decoupled from the quarks since quark loops are suppressed, giving rise to an emergent center symmetry in the sense defined in this paper: There exists a sector of nontrivial operators---in this case single-color-trace operators containing only gluon fields---in the theory whose correlation functions match,up to corrections which vanish as $1/N_c$, the correlation functions of a distinct theory---in this case, pure Yang-Mills theories--- which is explicitly invariant under center transformations.   
 
 Thus in  both  $2+1$ and $3+1$ space-time dimensions SU($N_c$) gauge theories with quarks in the fundamental representation both have Hagedorn spectra and an emergent center symmetry in conformity with the conjecture.
 
\subsection{SU($N_c$) Yang-Mills theories}
 
 In this subsection we show that SU($N_c$) theories in both 2+1 and 3+1 space time dimensions have Hagedorn spectra; since these theories have explicit center symmetry they are consistent with the conjecture.  To do so, we need to establish that the four conditions in Sect. \ref{Hagcond} are satisfied.  Condition \ref{cond1} is satisfied for glueballs by the standard reasoning of large $N_c$ QCD \cite{Witten:1979kh};  condition \ref{cond3} is known to be satisfied for these theories.   Moreover, as noted previously the planarity of large $N_c$ QCD ensures that condition \ref{cond4} is satisfied.  Thus to establish a Hagedorn spectrum, it is only necessary to establish condition \ref{cond2}. 
 
It is not necessary to consider pure Yang-Mills in 1+1 space time dimensions.  Gluons in 1+1 space time dimensions are not dynamical degrees of freedom and pure gluodynamics is trivial.  This is in sharp contrast to 1+1 dimensional theories containing matter field  in which case the gluon can act to mediate interactions between the matter.
 
 \subsubsection{2+1 space-time dimensions \label{YM2p1}}
 
 We begin by constructing a set of scalar operators for the case of 2+1 space-time dimensions to demonstrate that condition \ref{cond2} is satisfied. To do so we introduce the following building blocks :
\bea
O_0=F^{\alpha\beta}F_{\alpha\beta}F^{\alpha'\beta'}F_{\alpha'\beta'}F^{\alpha''\beta''}F_{\alpha''\beta''}\nn\\
O_1=(\epsilon^{\rho\mu\nu}F_{\mu\nu}\epsilon^{\sigma\mu'\nu'}F_{\mu'\nu'}\epsilon^{\lambda\mu''\nu''}
F_{\mu''\nu''}\epsilon_{\rho\sigma\lambda})^2 . \label{2p1}
\eea
The operators $O_0$ and $O_1$ have both a naive mass dimension of 9. Note that, as written $F_{\alpha\beta}$ is a matrix in color space: $F_{\alpha\beta}^a \lambda^a$ where $\lambda^a$ are the SU($N_c$) Gell-Mann matrices and the summation over $a$ is implicit. Since these operators contain explicit colors, they are not gauge invariant. However, tracing over the color of a product of these operators leads to gauge-invariant scalar operators. Let us denote an operator of this sort with a naive mass dimension of $d$, $J^d_b$, where $b$ is a one-dimensional array of $d/9$ numbers each of which can take the values 0 and 1, corresponding to operators 0 and 1:
\begin{equation}
J^d_{\{n_1, n_2, \cdots , n_{d/9} \} } \equiv {\rm Tr} \left ( O_{n_1} O_{n_2} \cdots O_{n_{d/9}} \right ) \; . \label{Jdef2}
\end{equation}
Thus for example $J^{45}_{\{1, 1, 0 , 1,0 \} }= {\rm Tr} \left ( O_1 O_1 O_0 O_1 O_0\right )$ and $J^{27}_{\{0, 1, 0  \} }= {\rm Tr} \left ( O_0 O_1 O_0 \right )$.

The operators defined in Eq~(\ref{Jdef2})  have a number of important properties. Firstly, by construction, they are local, gauge-invariant and have a  single-color trace. This is critical given condition \ref{cond2}. Secondly, the operators have the property
\begin{equation}
J^d_{\{n_1, n_2, \cdots , n_{d/9} \} } = J^d_{\{n_2, n_3 \cdots , n_{d/9}, n_1 \} }  \; ;
\end{equation}
this property follows from the cylic property of the trace. A final property is that at large $N_c$,  the operators $J^d_{b_1} ,J^d_{b_2}  ,J^d_{b_3} \cdots \}$ are linearly independent provided that  none of the $b$'s can be turned into another by cyclic permutation.  This is a consequence of the analysis of ref.~\cite{Cohen:2011yx}.

Given these properties, it is straightforward to establish condition \ref{cond2}.    Let us first focus on cases where $d$ is a multiple of 9.   Denote as $N_d$ the total number of linearly independent single-trace gauge-invariant scalar operator with mass dimension less than $d$  and  denote as $N^d_J$, the number of linearly independent operators of the type $J^d_b$ defined in Eq.~(\ref{Jdef2}). By construction $N^d_J$ is the number of distinct sequences of $d/9$ bits with cyclic permutations identified and, as such, it follows from elementary consideration of number theory and combinatorics\cite{Cohen:2009wq} that 
\begin{equation} N_d^J \ge \frac{2^{\frac{d}{9}}+2(\frac{d}{9}-1)}{\frac{d}{9}} \label{number} \end{equation}
with the equality being saturated, if and only if, $d$ is prime.  Since by construction $N^d \ge N_d^J$, it follows that for sufficiently  large $d$, 
\begin{equation} N_d \ge \exp \left ( \left (\frac{\log(2)}{9} -\epsilon \right ) d \right ) \end{equation} 
for any positive $\epsilon$. This establishes the exponential growth needed and thus, condition \ref{cond2} is satisfied and if the assumptions noted above hold, a Hagedorn spectrum is established.  

\subsubsection {3+1 space-time dimensions}

The case of 3+1 space-time dimensions is formally quite similar. For simplicity here we will look at both scalar and pseudoscalar operators together.  We again start by introducing building blocks:
\bea
O_0=F^2\nn\\
O_1=F\tilde{F}\nn \label{3p1}\\ 
\eea
which again are matrices in color space.  Note  that $O_0$ is scalar $O_1$ is pseudoscalar.  Both of these operators have naive mass dimension  4.  We again use an analogous construction in  Eq.~(\ref{Jdef2}) to produce local gauge-invariant single-color trace operators:
\begin{equation}
J^d_{\{n_1, n_2, \cdots , n_{d/9} \} } \equiv {\rm Tr} \left ( O_{n_1} O_{n_2} \cdots O_{n_{d/4}} \right ) \; . \label{Jdef3}
\end{equation}
 The only difference from the 2+1 space-time dimensional case is the different choices for operators $O_0$ and $O_1$ and the corresponding differences in naive mass dimension. Again the operators \{$J^d_{b_1} ,J^d_{b_2}  ,J^d_{b_3} \cdots \}$ are linearly independent provided that  none of the $b$'s can be turned into another by cyclic permutation.   Thus, by precisely the same type of reasoning  as in the previous section, it follows that for sufficiently  large $d$, 
\begin{equation} N_d \ge \exp \left ( \left (\frac{\log(2)}{4} -\epsilon \right ) d \right )  \, \end{equation}  
and condition \ref{cond2} is satisfied.  Again, this demonstrates a Hagedorn spectrum at large $N_c$ given our assumptions

 \subsection{SU($N_c$) gauge theories with quarks in the adjoint representation}

 SU($N_c$) gauge theory with quarks in the adjoint representation has an explicit center symmetry. In this subsection it will be shown that SU($N_c$) gauge theories with either massive or massless adjoint quarks in 3+1 and 2+1 space-time dimensions and only massive adjoint quarks in 1+1 space-time dimensions exhibit Hagedorn spectrum. These theories also happen to be the ones where the explicit center is not broken spontaneously. We emphasize again that in SU($N_c$) gauge theories with massless adjoint quarks in 1+1 space-time dimensions center symmetry is spontaneously broken and there is no Hagedorn spectrum either as discussed in Sec. \ref{massless}. This is expected assuming the conjecture at the heart of this paper is correct.
 
 Two dimensional QCD with massive adjoint quarks is similar to pure Yang-Mills theory in four dimensions. Upon dimensional reduction the four dimensional theory becomes a two dimensional theory coupled to adjoint scalars. This is similar to the theory which is obtained by bosonizing the adjoint fermions. The subtle issue, that we had already discussed earlier, is that the string tension in two dimensional QCD is proportional to the quark mass. The theory with massive quarks is expected to exhibit a Hagedorn spectrum and evidence for that had been given in various DLCQ simulations \cite{Bhanot:1993xp,Demeterfi:1993rs,Gross:1997mx}.
 
The demonstration that these theories have  Hagedorn spectra amounts to a demonstration that condition \ref{cond2} is satisfied for them.  For the case of $2+1$ and $3+1$ dimensions this is trivial given the analysis of the pure Yang-Mills theories: one can simply use precisely the operators defined in Eq~(\ref{Jdef2}) and Eq~(\ref{Jdef2}) using the operators given in Eqs.~(\ref{2p1}) and (\ref{3p1}) for 2+1 and 3+1 space-time dimensions respectively.  These were already shown to satisfy condition \ref{cond2}.  As it happens,  in these theories there is a much larger set of local gauge-invariant single-color trace  operators with a given naive mass dimension  than in pure Yang-Mills theories, since one can build operators including quark-antiquark pairs, but one need not exploit this fact to demonstrate a Hagedorn spectrum at large $N_c$.

For 1+1 space-time dimensions,  one needs  quark-antiquark operators to proceed. We introduce two new operators, 
\bea
O_0 = \bar{q} q \nn \\
O_1 = \bar{q} \gamma_5 q \; . \label{1p1ops}
\eea
Note that since these quarks are in the adjoint representations, these operators, like the operators defined in  Eqs.~(\ref{2p1}) and (\ref{3p1})  are matrices in color space. The operators are of naive mass dimension 1. One can again exploit the structure of Eq~(\ref{Jdef2}) to produce a set of local, gauge-invariant single-color-trace scalar and pseudoscalr operators: 
\begin{equation}
J^d_{\{n_1, n_2, \cdots , n_{d/9} \} } \equiv {\rm Tr} \left ( O_{n_1} O_{n_2} \cdots O_{n_{d}} \right ) \; . \label{Jdef1}
\end{equation}  Again the operators \{$J^d_{b_1} ,J^d_{b_2}, J^d_{b_3} \cdots \}$ are linearly independent provided that  none of the $b$'s can be turned into another by cyclic permutation. Thus, it follows that for sufficiently large $d$, 
\begin{equation} N_d \ge \exp \left ( \left ({\log(2)} -\epsilon \right ) d \right )  \ , \end{equation}  condition \ref{cond2} is satisfied and, given our assumptions, a Hagedorn spectrum exists at large $N_c$.

 \subsection{SU($N_c$) gauge theories with quarks in two-index representations}
 
SU($N_c$) gauge theories can have quarks in two-index symmetric and anti-symmetric representations.   In such theories each quark is labeled by two fundamental color indices  with $q_{ab}= - q_{ba}$ and $q_{ab}= q_{ba}$ for  antisymmetric and symmetric representations respectively.   Such theories are renormalizable and  are asymptotically free in 3+1 space-time dimensions or fewer, provided that there are not too many flavors. There has been significant interest in such theories since 2003, when it was argued  that a sector of the theory becomes equivalent at large $N_c$ to an analogous sector of theories with quarks in the adjoint representation\cite{Armoni:2003gp,Armoni:2004ub}. This is significant for the present purpose since, the theories with matter in the adjoint representation have an explicit center symmetry, so that these theories have an emergent center symmetry at large $N_c$ in the sense used in this paper. Note moreover, that the argument for large $N_c$ equivalence between theories with quarks in the two-index representations and quarks in the adjoint representation does not depend on the space-time dimension

The equivalence between these two theories can be easily demonstrated at the perturbative level. The full non-perturbative proof in favor of the equivalence is more challenging \cite{Armoni:2003gp,Armoni:2004ub}. It was shown  \cite{Unsal:2006pj} in that a necessary and sufficient condition for the equivalence to hold is that charge conjugation symmetry is not broken in the theory with antisymmetric quarks. We assume that this is indeed the case.
%We note that while there is a simple perturbative argument suggesting why the equivalence might hold and additional non perturbative arguments \cite{Armoni:2003gp}, there  has been some controversy as to whether the large $N_c$ equivalence between sectors of the theories with  quarks in the two-index representation and the adjoint representation has been fully proved in a nonperturbative fashion. It has been argued  that the demonstration depends on  an additional assumption that a discrete symmetry remains unbroken \cite{Unsal:2006pj}. It is noted moreover that the symmetry is not guaranteed to be unbroken and is known to break for a certain type of compactification of the space-time. Despite this concern, In this paper, we will take it as being at least  highly plausible that the equivalence does hold for the case on an uncompactified space-time in 1+1, 2+1 and 3+1 space-time dimensions.

If the conjecture of this paper is correct, and these theories have an emergent center, then they should have a Hagedorn spectrum at large $N_c$. This is easy to establish given the assumptions made earlier. Again, conditions \ref{cond1}, \ref{cond3} and \ref{cond4} should hold. Thus, the issue comes down to condition \ref{cond2}. For the case of $2+1$ and $3+1$ dimensions this is again trivial given the previous analysis: One can, again simply use precisely the operators defined in Eq~(\ref{Jdef2}) and Eq~(\ref{Jdef2}) using the operators given in Eqs.~(\ref{2p1}) and (\ref{3p1}) for 2+1 and 3+1 space-time dimensions respectively which are known to satisfy condition \ref{cond2}.  The case of 1+1 dimensions is also straightforward.  One again begins with the operators in Eq.~(\ref{1p1ops}).  The key thing is that operator $\bar{q}  q$ is to be interpreted as a matrix in color space with matrix elements $\left(\bar{q}  q\right)^a_b = \sum_c \bar{q}^{a c}q_{c b}$  and analogously for $\bar{q}  \gamma_5 q$.  Given this matrix structure, one can then use Eq.~(\ref{Jdef1}) to define operators which, as has already been shown satisfy condition {\ref{cond2}). Thus, if the assumptions we are using are correct, the theory has a Hagedorn spectrum at large $N_c$.

The bottom line here is that it is highly plausible that SU($N_c$) gauge theories with quarks in two-index representations, in 1+1, 2+1 or 3+1 space-time dimensions have both an emergent center symmetry and hagedorn spectra. This is as expected from the conjecture.

\subsection{SU($N_c$) theories with quarks in multiple representations}

The analysis in the previous subsections is rather robust and continues to hold even if multiple types of matter fields are included. Thus, for example, consider a theory with one flavor of two-index quark---a theory for which the conditions for a Hagedorn spectrum at large $N_c$ are satisfied and for which there is a very plausible argument that the theory has an emergent center symmetry---and then add to it one flavor of fundamental quark. It should be clear, that the conditions for a Hagedorn spectrum remain satisfied at large $N_c$ and the argument for an emergent center symmetry is also unaffected. Indeed, quite generally,  if the additional fields do not alter conditions for a Hagedorn spectrum (by, for example, destroying asymptotic freedom), one expects theories of this type to both have a Hagedorn spectrum and an emergent center symmetry.

\subsection{SO($N_c$) and Sp($N_c$)  theories}

It has long been believed that $SU(N_c)$ gauge theories were equivalent (within appropriate sectors) at large $N_c$ to gauge theories based on orthogonal and symplectic groups \cite{Lovelace:1982hz}.    A modern perspective on this is that SO(2$N_c$) or Sp(2$N_c$) theories are related by an orbifold equivalence to SU($N_c$) theories \cite{Hanada:2011ju,Cherman:2010jj}.

For example it  is known that $SU(N_c)$ gauge theory with $N_f$ fundamental fermions is a daughter theory of $SO(2N_c)$ or $Sp(2N_c)$ where the two gauge theories are related by orbifold equivalence \cite{Hanada:2011ju}, \cite{Cherman:2010jj}. All 
correlation functions belonging to the neutral sector of the two theories coincide to leading order in $N_c$ provided the symmetry involved in the projection is not spontaneously broken. Again we take it to be plausible that the projection symmetry is not broken and the equivalence holds enabling us to extend our arguments and conclusions in all the previous examples while replacing $SU(N_c)$ by $Sp(2N_c)$ and $SO(2N_c)$. Because the correlation functions match those of SU($N_c$) theories, all of the previous examples in which it is shown that the theories have Hagedorn spectra at large $N_c$ go through. Similarly, all of these theories have an emergent center at large $N_c$ since the analogous SU($N_c$) theories do.

We note in passing that the argument based on orbifold projections can only be formulated for orthogonal groups with even $N_c$. However, we take it as highly plausible that a large $N_c$ limit exists for the observables of interest for theories based on SO($N_c$). If this is the case the difference between theories with $N_c$ even and odd is a $1/N_c$ correction and vanishes at large $N_c$.

\section{Discussion}\label{sec:conclusion}

In this paper, we have argued that the existence of a Hagedorn spectrum at large $N_c$ should be regarded as an indicator of confinement in much the same way that the vanishing of the Polyakov loop or the area law for the Wilson loop are indicators of confinement.  As with these other indicators of confinement, it does not apply to all theories, but requires the theory to be center symmetric. There is a subtlety in that the theory need not have an explicit center symmetry; an emergent center symmetry at large $N_c$ is sufficient.

Substantial evidence for this conjecture has been presented in this paper, although it is not definitive. In effect what has been shown is that large classes of theories exist for which there is both a very strong evidence that they have Hagedorn spectra at large $N_c$ and which are either explicitly center symmetric or for which there are strong plausibility arguments that they acquire an emergent center symmetry at large $N_c$. This evidence, by itself is not completely compelling. One limitation, of course is that we have used plausibility arguments as opposed to rigorous mathematical proofs. While this is unfortunate, it is the best one can do given the state of the art. There is, however, another potential issue which might weaken the plausibility of the conjecture: it seems that both Hagedorn spectra and emergent center symmetry are almost ubiquitous at large $N_c$. Thus one could imagine that virtually every confining gauge  theory has a Hagedorn spectrum at large $N_c$ and, for unrelated reasons, virtually every confining gauge theory has an explicit or emergent center symmetry at large $N_c$. Fortunately, there is a reason to discount this last concern. Note that Hagedorn spectra and center symmetry at large $N_c$ are only {\it almost} ubiquitous. There are a few situations in which a Hagedorn spectrum does not emerge at large $N_c$. In those situations either there is not an emergent center---as happens in the 't Hooft model and the Veneziano limit---or the theory is known to be a conformal as opposed to confining phase.

In the introduction, the possibility of a connection between  Hagedorn spectra and confinement was motivated by the fact that confinement is often associated with flux tubes which for high-lying states at large $N_c$ act like unbreakable strings and that simple string theories based on unbreakable strings give rise to Hagedorn spectra due to vibrations in transverse directions. To the extent that confinement is associated with center symmetry\cite{Cohen:2014swa, Greensite:2011zz}, one might naturally expect that at large $N_c$ the Hagedorn spectrum and center symmetry are connected. It is worth remarking that while this may well be part of what is happening, it cannot the entire story.  Note that theories in 1+1 dimensions that have massive quarks in either the adjoint representation or two-index representation are expected to have  Hagedorn spectra and either explicit or emergent center symmetry (for the cases of quarks in the adjoint or two index quarks respectively).  However, in 1+1 dimensions flux tubes cannot vibrate in transverse directions leading to a Hagedorn spectrum for the simple reason that transverse directions do not exist. Thus the connection between Hagedorn spectra and center symmetry is more profound.  
%  Thus, a Hagedorn spectrum appears to exist even when the dynamics is not stringy.  Note however, that even in this non-stringy context, the existence of  a Hagedorn spectrum appears to go along with center symmetric theories.  This further suggests that there may be a deep connection between Hagedorn spectra at large $N_c$ and center symmetry.

A final point worth clarifying illustrates the nature of the emergent center symmetries and involves the case  of SU($N_c$) gauge theory with matter fields  in the fundamental representation. Note that the emergent center in this case follows from the fact that at large $N_c$ the gluodynamics does not depend on the dynamics of the  quarks.  This clearly satisfies the definition of ``emergent symmetry'' introduced in this paper; there exists a sector of the theory (in this case operators built from gluons) whose correlation functions at large $N_c$ match those of a theory with an explicit center (in this case pure Yang-Mills theory).   Given that the emergent center is the same as in pure Yang-Mills, it is not surprising that glueballs in the theory have a Hagedorn spectrum as they do in pure Yang-Mills.  However, the mesons in the theory also have a Hagedorn spectrum\cite{Cohen:2011yx} and clearly are not associated with states in Yang-Mills theory.  The class of states with a Hagedorn spectrum  includes those not in the common sector. In some sense this should not be too surprising. Note that the emergent center is associated with gluodynamics. When one looks at the mesons, the states  obviously contain quarks. However, the highly excited states associated with the Hagedorn behavior are gluonic type excitations. That is, in the language of the quark model, these states are ``hybrids'' rather than pure mesons and it is the excitation of the gluonic degrees of freedom which gives rise to the Hagedorn spectrum  for mesons. This can be understood in a number of ways. If one thinks about a flux tube picture, these states are modeled by a flux tube with quarks at the ends. The states are vibrational excitations of the flux tube---{\it i.e.} of the gluons in the theory. Similarly if one uses the argument of ref.~ \cite{Cohen:2011yx} and looks at the structure of the operators  with meson quantum numbers, it is the inclusion of gluonic structures between the quarks which allows the number of operators to grow exponentially with dimension.  The fact that spectrum assumes a Hagedorn structure even outside the common sector in such theories suggests that the key point in the connection between a Hagedon spectrum and an emergent center symmetry, is simply the {\it existence} of the emergent center regardless of its details.  
\section*{Acknowledgements}

The work of TDC and SS is supported by the U.S. Department of Energy through grant number DEFG02-93ER-40762. AA is grateful to the U.K.~Science and Technology Facilities Council (STFC) for financial support under grants ST/J000043/1 and ST/L000369/1.  The authors thank A. Cherman for insightful conversations.

\bibliographystyle{unsrt}
\bibliography{hagedorn4}

\end{document}